\documentstyle[12pt,prl,aps,psfig]{revtex}

\begin{document}

\renewcommand{\baselinestretch}{1}

\vskip 0.9truecm

\title{{\flushright{\small CTP-2536\\Imperial/TP/95-96/50\\PUPT-1624\\
hep-th/9605169\\}}\vskip 0.5truecm
\Large\bf
The Ising model with a boundary magnetic field\\ on a random surface}
\author{Sean M.  Carroll\cite{sa}}
\address{Center for Theoretical Physics, Laboratory for Nuclear
Science and Department of Physics,\\
Massachusetts Institute of
Technology, Cambridge MA 02139}
\author{Miguel E.  Ortiz\cite{ma}}
\address{Blackett Laboratory, Imperial College of Science, Technology
and Medicine,\\ Prince Consort Road, London SW7 2BZ, UK}
\author{Washington Taylor IV\cite{wa}}
\address{Department of Physics, Princeton University,
Princeton, New Jersey 08544}

\maketitle
\vskip 1truecm

\begin{abstract}
The bulk and boundary magnetizations are calculated for the critical
Ising model on a randomly triangulated disk in the presence of a
boundary magnetic field $h$.  In the continuum limit this model
corresponds to a $c = 1/2$ conformal field theory coupled to 2D
quantum gravity, with a boundary term breaking conformal invariance.
It is found that as $h$ increases, the average magnetization of a bulk
spin decreases, an effect that is explained in terms of fluctuations
of the geometry.  By introducing an $h$-dependent rescaling factor,
the disk partition function and bulk magnetization can be expressed as
functions of an effective boundary length and bulk area with no
further dependence on $h$, except that the bulk magnetization is
discontinuous and vanishes at $h = 0$.  These results suggest that
just as in flat space, the boundary field generates a renormalization
group flow towards $h = \infty$.  An exact analytic expression for the
boundary magnetization as a function of $h$ is linear near $h = 0$,
leading to a finite nonzero magnetic susceptibility at the critical
temperature.
\end{abstract}

\medskip

%\narrowtext
%\twocolumn
\newpage

\renewcommand{\baselinestretch}{1}
\baselineskip 16pt

The Ising model with a boundary magnetic field\cite{mw} has been of
renewed interest recently as a simple example of a two-dimensional
integrable field theory with nontrivial boundary interactions
\cite{gz,cz,klm}.  The only boundary conditions for the Ising model which
preserve conformal invariance\cite{cardy} are free boundary conditions
(where the boundary field $h$ vanishes), and fixed spin boundary
conditions (where $h = \pm \infty$).  Putting an arbitrary field $h$
on the boundary generates a renormalization group (RG) flow which
goes away from the free boundary condition towards the fixed boundary
condition \cite{al}.

Another subject of recent interest has been the effect of different
boundary conditions in string theory\cite{polchinski}.  Just as there
are two types of conformally invariant boundary conditions for the
Ising theory, a conformal field theory of a single bosonic field can
have two conformally invariant boundary conditions: Neumann and
Dirichlet.  By considering the continuum limit of the Ising model as a
single free fermion, in the context of superconformal field theory it
can be shown that boundary conditions in these two models are related
by supersymmetry, with Neumann corresponding to free Ising spins and
Dirichlet corresponding to fixed spins.

In this letter we consider the effect of a boundary magnetic field on
the Ising model on a random surface (the noncritical string with $c =
1/2$).  This theory describes a single Ising spin (or equivalently a
free fermion) coupled to 2D quantum gravity.  
%The partition function
%is computed by integrating $Z_g e^{-\mu A}$ over distinct metrics $g$
%on a surface $\Sigma$ ({\it i.e.}, metrics which are not related by
%diffeomorphisms), where $Z_g$ is the partition function of the
%conformal field theory on $\Sigma$ with metric $g$, $A$ is the area of
%$\Sigma$, and $\mu$ is a cosmological constant.
%
The Ising model on a random surface can be studied in several ways.
One approach is to use the continuum formulation of noncritical string
theory as a conformal field theory coupled to a Liouville
field\cite{ddk,kpz}.  Another approach is to describe the model as a
matrix model, involving a sum over discrete surfaces\cite{kazakov}; in
the discrete formalism, a continuum limit can be taken by tuning
coupling constants until the surfaces become arbitrarily large; in
this limit the theory corresponds to the continuous Liouville theory.
In this work we will use a discrete formulation of the Ising model on
a random surface; to calculate correlation functions in this theory we
use the method of discrete loop equations developed in two previous
papers \cite{cot1,cot2}.  Similar methods were discussed
in\cite{staudacher,dl}.

We present here only the results of our investigation.  The details of
the calculations, which are algebraically tedious, will be
presented in a later publication.

In the discrete formulation, the Ising model on a random surface of
disk topology has a partition function which is given by a sum over
all possible triangulations of the disk.  For each triangulation, the
Boltzmann weight is given by placing a single Ising spin on each
triangle, and summing over all possible spin configurations, giving
the Ising partition function on that particular geometry.
This model can be written as a matrix model\cite{kazakov}
\begin{equation}
{\cal Z} (g, c)= \int {\rm D} U \; {\rm D} V \;
\exp \left(-NS (U,V) \right)\ ,
\label{eq:matrixising}
\end{equation}
with
\begin{equation}
S (U,V)={\rm Tr}\left[\frac{1}{2} (U^2 + V^2)-cUV -
\frac{g}{3} (U^3+V^3)\right]\,,
% \label{eq:}
\end{equation}
where $U$ and $V$ are $N\times N$ hermitian matrices.  

The first calculation we wish to consider is that of the disk
amplitude when the spins on the boundary are subjected to an external
magnetic field $h$.  In
the matrix model language, dropping all factors of $N$ as we work
in the large $N$ limit, we wish to compute
\begin{equation}
\phi (h, x, g, c) =
\sum_{k = 0}^{\infty}  x^k  \langle
{\rm Tr}\; (e^{h}U + e^{-h}V)^k \rangle
\equiv \sum_{k = 0}^{\infty} x^kp_k\ ,
\label{eq:phi}
\end{equation}
where $p_k=\langle {\rm Tr}\; (e^{h}U + e^{-h}V)^k \rangle$ is the
amplitude for a disk with $k$ boundary spins subject to the boundary
field $h$.  A method for calculating such amplitudes was described
in\cite{cot1}.  This yields a quartic equation satisfied by $\phi$, in
which the coefficients are functions of $g, c, h, x,$ and $p_i$ with
$i < 4$.  We omit this equation for space considerations.

The quartic equation for $\phi$ gives an exact algebraic solution for
the disk partition function of the discrete theory.  To find the
solution in the continuum limit, we must find the critical values for
$x$ and $g$ at which $\phi$ approaches a singular point.  
The Ising model is critical for the coupling
$c_c=(-1 + 2 \sqrt{7})/27$.  The critical
value for $g$ is known to be \cite{kazakov} $g_c=
\sqrt{10 c_c^3}$.  
After an analysis of the critical behavior of $\phi$, we find
\begin{equation}
%x_c =  \frac{g_c(1 + 2 \sqrt{7})w^3}{1 + w^2 (\sqrt{7}-1) + w^4 (2 +
%\sqrt{7})}
x_c = \frac{g_c(1 + 2 \sqrt{7})e^{h}}{ e^{-2h} + (\sqrt{7}-1) + e^{2h}
(2 +\sqrt{7})}\ .
% \label{eq:}
\end{equation}
This expression is only valid for $h \geq 0$; $x_c$ is nonanalytic at
$h = 0$.  Throughout the remainder of this letter we will restrict
attention to the case $h   \geq  0$; related expressions arise when $h <
0$.

To take the continuum limit, we expand around the critical values
$g = g_c e^{-\epsilon^2 t}$, 
$x = x_c e^{-\epsilon z}$.
Expanding $\phi$ in $\epsilon$, we find
\begin{equation}
\phi =  \phi_a (z, t, \epsilon)
+ \frac{\epsilon^{4/3}}{5\cdot
2^{7/3}\alpha (h)}  \Phi
(Z, T)+{\cal O} (\epsilon^{5/3})\ ,
\label{eq:phiuniversal}
\end{equation}
where $\phi_a$ is
analytic in $\epsilon$.
The second term is nonanalytic and
describes the behavior in the continuum limit.  Here
\begin{equation}
\Phi\equiv
\left(Z+\sqrt{Z^2-4T}\right)^{4/ 3}
+\left(Z-\sqrt{Z^2-4T}\right)^{4/ 3}\ ,
\label{bigphi}
\end{equation}
$t$ is rescaled by a constant factor $t=T/5$, and
$z$ is rescaled by an $h$-dependent factor, $z = \alpha (h)Z$
where for $h > 0$
\begin{equation}
\alpha (h) = \frac{(1 + e^{2h})}{e^{-2h} + 
(-1+\sqrt{7}) + e^{2h} (2 + \sqrt{7})}\ .  
\label{eq:a}
\end{equation}
At $h = 0$, the scaling factor $\alpha$ is discontinuous and goes to
$\alpha (0) =1/(\sqrt{2} + \sqrt{14})$; the constant factor in the
universal term in (\ref{eq:phiuniversal}) also changes discontinuously
at this point.  Note that the specific form (\ref{eq:a}) for $\alpha$
depends upon the discretization we have chosen for random surfaces.

The universal term in (\ref{eq:phiuniversal}) can be converted into
the asymptotic form of the disk amplitude $ \tilde{\phi} (l,a)$
for fixed boundary length
$l$ and disk area $a$.  These forms of the amplitude are related
through a Laplace transform
\begin{equation}
\phi_{(4/3)} (z, t) =  \int {\rm d} l \int {\rm d} a \; \;e^{-zl-ta}
\tilde{\phi} (l, a)\ .
% \label{eq:}
\end{equation}
Inverting the Laplace transform, we have
\begin{equation}
\tilde{\phi} = 
{1\over 25\sqrt{3} \pi} L^{1/3} A^{-7/3} e^{-L^2/A}\ ,
% \label{eq:}
\end{equation}
with the rescalings $L = \alpha (h) l$, $A = {a/5}$.
Up to an irrelevant multiplicative constant, this is precisely the
form of the disk amplitude when the boundary conditions are conformal
\cite{mss,staudacher,gn,cot2} (i.e., with $h = 0$ or $h = \pm
\infty$); however, the boundary length $l$ is rescaled by the factor $
\alpha (h)$ which depends on the boundary magnetic field.  

The boundary magnetization for a spin on the boundary of a disk with
$k$ boundary edges and $n$ triangles is given by 
\begin{equation}
\langle m \rangle =
{\langle{\rm Tr}(e^h U - e^{-h}V)
(e^hU+e^{-h}V)^{k-1}\rangle_n\over\langle{\rm Tr}
(e^hU+e^{-h}V)^{k}\rangle_n}\ ,
% \label{eq:}
\end{equation}
where by $\langle \rangle_n$ we indicate a sum over triangulations
restricted to geometries with $n$ spins (the coefficient of $g^n$ in
an expansion in $g$).  We therefore look at the expectation value of
the spin at a marked point on the boundary, that is
\begin{equation}
\psi=\sum_{k =0}^\infty x^{k+1} \langle{\rm Tr}(e^h U -
e^{-h}V) (e^hU+e^{-h}V)^{k}\rangle\ .
% \label{eq:}
\end{equation}
When $h = 0$, $\psi$ vanishes by symmetry.  When $h \neq 0$, we can
compute $\psi(h)$ by the method of loop equations, giving an
equation relating $\psi(h)$ to $\phi(h)$.  Solving this equation, we
can find the critical expansion of $\psi(h)$ about the critical point
and the inverse Laplace transform $\tilde{\psi}$ of the universal part
of $\psi$.  The boundary magnetization in the continuum limit is then
given by (for $h > 0$)
\begin{equation}
\langle m \rangle={\tilde{\psi}\over\tilde{\phi}}=
{(e^{h} -e^{-h})(e^{-h} + (2+\sqrt{7})e^{h})\over
(e^{-2h} +(-1+\sqrt{7}) + (2+\sqrt{7})e^{2h})}\ .
\label{eq:boundarymagnetization}
\end{equation}
Note that the boundary
magnetization is
independent of $l$ and
$a$.

A graph of the boundary magnetization is shown in Fig.~1 (bold
curve).  As expected, with no field the magnetization is zero,
and for an infinite field the magnetization is 1.  This
result is compared with the boundary magnetization in flat space,
computed by McCoy and Wu\cite{mw} (dashed curve).  
Whereas in flat
space the magnetization scales as $h \ln h$  for small $h$, leading to
a divergence in the magnetic susceptibility at the critical
temperature,  on a random surface we find that the magnetization is
linear at $h = 0$, with a finite susceptibility
\begin{equation}
\chi =\partial_h \langle m \rangle |_{h = 0} = \frac{1 + 2
\sqrt{7}}{3} \ .
% \label{eq:}
\end{equation}
The two point boundary magnetization
can be computed in a similar way, and is found to be equal to the
square of the one point
magnetization.  

\begin{figure}
\vskip -3.75cm
\centerline{
\psfig{figure=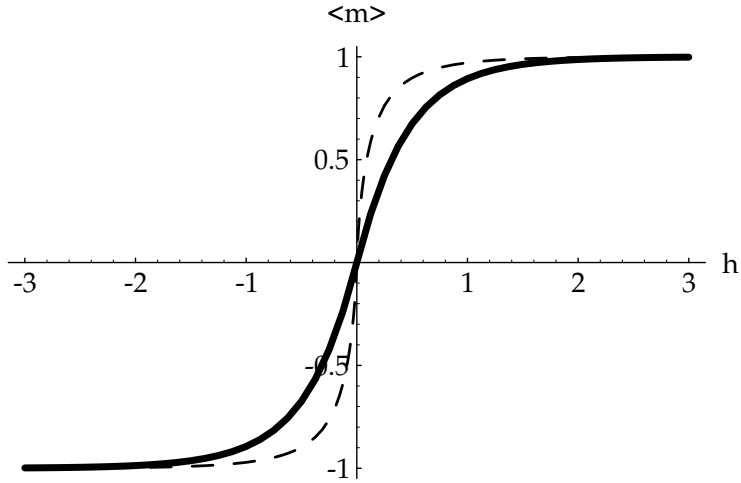,angle=0,height=13.5cm}}
\vskip -3cm
\caption{Boundary magnetization $\langle m \rangle$ as a function of
boundary field $h$ in flat space (dotted line) and on a random surface
(bold line)}
\end{figure}

Consider now the average bulk magnetization with a boundary magnetic
field, on a disk with boundary length $k$ and area $n$.
\begin{equation}
\langle M\rangle =
{\langle{\rm Tr}
(e^hU+e^{-h}V)^{k}{\rm Tr}(U-V)\rangle_n\over\langle{\rm Tr}
(e^hU+e^{-h}V)^{k}{\rm Tr}(U+V)\rangle_n}\ .
% \label{eq:}
\end{equation}
This can be evaluated by considering cylinder amplitudes with one
boundary having a boundary magnetic field, and the other with a single
boundary edge.  The second boundary represents a marked point on the
bulk.  Again such a quantity can be computed by the method of loop
equations\cite{cot2}.  After some algebra, it can be shown that the
bulk magnetization in the continuum limit is given by the simple
expression
\begin{equation}
\langle M \rangle=
L^{1/3}A^{-1/3}\ .
\label{eq:bulkmagnetization}
\end{equation}
Since $l$ and $a$ are measured in lattice units, in the continuum
limit $L \ll A$ so the magnetization is always less than 1.  More
precisely, in the continuum limit, $\langle M \rangle$ scales as
$\delta^{1/3}$ where $\delta$ is the lattice spacing; this agrees with
the scaling dimension of the gravitationally dressed spin
operator\cite{ddk,kpz}.  Note that this form of the magnetization is
independent of $h$ except for the dependence through the scaling
factor $\alpha$ incorporated in $L$.  At $h = 0$, this magnetization
is discontinuous and vanishes.

We have found that both the disk partition function and the bulk
magnetization are naturally expressed in terms of a rescaled boundary
length $L = \alpha (h)l$.  
%An interesting feature of the bulk
%magnetization (\ref{eq:bulkmagnetization}) is that when expressed in
%terms of the actual boundary length $l$, the magnetization is a
%function which for fixed values of $l$ and $a$ {\em decreases} as the
%boundary magnetic field $h$ increases.
An interesting feature of the bulk magnetization
(\ref{eq:bulkmagnetization}) is that, with the particular choice of
discretization scheme we have used here, when expressed in terms of
the actual boundary length $l$, the magnetization is a function which
for fixed values of $l$ and $a$ {\em decreases} as the boundary
magnetic field $h$ increases.
This counterintuitive result
can be explained in terms of the influence of the boundary field on
the disk geometry.  In the vicinity of the disk boundary, the
existence of a boundary field causes a local fluctuation of the
discrete geometry which depends upon the magnitude of the boundary
field.  In the continuum limit, this effect is restricted to a
vanishingly small region near the boundary.
%, since the effective field
%in the bulk of the disk scales as $d^{-1/3}$ where $d$ is the distance
%from the boundary as measured in lattice units.  
The significance of the rescaled boundary length $L$ is that the
effects of the boundary magnetic field can be described by using in
place of the boundary length $l$ an effective boundary length $L =
\alpha (h)l$ (see Fig.~2).  In the bulk of the disk in the continuum
limit, for any nonzero boundary field $h$, all the physics is
identical to the physics which would occur on a disk of boundary
length $L/\alpha_\infty$ with infinite magnetic field on the boundary,
where $\alpha_\infty = 1/(2 + \sqrt{7})$ is the limit of the scaling
factor as $h \rightarrow \infty$.  Since $\alpha (h)$ is a decreasing
function of $h$, the effective boundary length $L$ decreases for fixed
$l$ as $h$ increases.  As in Fig.~2, a decrease in $L$ for fixed $A$
forces the disk to deform so that the spins move further away from the
boundary, causing a net decrease in the average magnetization.

\begin{figure}
\vskip -1.05cm
\centerline{
\psfig{figure=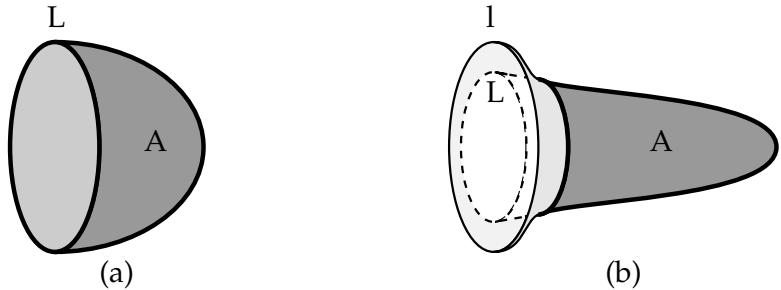,angle=0,height=6.75cm}}
\vskip -1.8cm
\caption{(a) A portion of a surface with area $A$ and boundary length
$L$, (b) A disk with boundary length $l$ and effective boundary length
$L < l$ is stretched so that an average point is further from the
boundary}
\end{figure}

Another interesting feature of the bulk magnetization is that when
expressed in terms of the rescaled boundary length $L$ it is
independent of $h$, except where $h = 0$, when the magnetization
vanishes.  This result indicates that just as in flat space, any
nonzero magnetic field produces a renormalization group flow whose
fixed point limit is the infinite magnetic field boundary condition.

Finally, it should be noted that of the results presented here, the
scaling factor $\alpha (h)$ in (\ref{eq:a}) and the explicit form of
the magnetization (\ref{eq:boundarymagnetization}) have a functional
dependence on $h$ which depends on the set of triangulations we have
used for the disk.  Just as in the case of the flat space Ising model,
different choices of lattice discretization will give rise to
different critical values $g_c$, different rescaling functions $\alpha
(h)$, and boundary magnetizations with different functional dependence
on $h$.  Certain of the results obtained here -- the functional
dependences of the disk partition function and the bulk magnetization
on the rescaled boundary length $L$ and area $A$, and the fact that
the boundary magnetic susceptibility is finite and nonzero -- should
be independent of the choice of discretization, however, and should
give the same continuum limit in any formulation of the theory.  On
the other hand, the result that the bulk magnetization decreases with
increasing boundary field is not necessarily a universal result, since
it depends upon the explicit form of the rescaling function $\alpha
(h)$.

\acknowledgments

We would like to acknowledge helpful conversations with D.\ Abraham,
V.\ Kazakov, and L.\ Thorlacius.  We are also grateful to the Someday
Cafe.  SC was supported in part by the National Science Foundation
under contract PHY92-00687 and in part by the U.S. Department of
Energy under cooperative agreement DE-FC02-94ER40818.  MO acknowledges
the financial support of the PPARC, UK, the European Community under a
Human Capital and Mobility grant, and the hospitality of MIT and
Utrecht University where parts of this work were carried out.  WT was
supported in part by the divisions of Applied Mathematics of the U.S.
Department of Energy (DOE) under contracts DE-FG02-88ER25065 and
DE-FG02-88ER25066, in part by the U.S.  Department of Energy (DOE)
under cooperative agreement DE-FC02-94ER40818, and in part by the
National Science Foundation (NSF) under contract PHY90-21984.

\renewcommand{\baselinestretch}{1}

\end{document}